\documentclass[onecollarge]{svjour2}        

\usepackage{graphicx}
\usepackage{epstopdf}
\usepackage{mathptmx}      	
\usepackage{amsfonts}
\usepackage{amsmath,amssymb,bm}
\usepackage[english]{babel}
\usepackage[utf8]{inputenc}
\usepackage[amssymb]{SIunits}

\journalname{Journal of Low Temperature Physics}

\begin{document}

\title{Self-similar expansion of the density profile in a turbulent Bose-Einstein condensate}

\author{M. Caracanhas \and
        A. L. Fetter \and
        S. R. Muniz \and
        K. M. F. Magalh\~aes \and
        G. Roati    \and
         G. Bagnato \and
        V. S. Bagnato
        }

\institute{
            M. Caracanhas \and
            S. R. Muniz \and
            K. M. F. Magalh\~aes \and
            G. Bagnato  \and
            V. S. Bagnato \at
                Instituto de Física de São Carlos, Universidade de São Paulo,\\
                Caixa Postal 369, 13560-970 São Carlos, SP, Brazil \\
               Tel.: +55-16-3371-2012\\
              \email{srmuniz@ifsc.usp.br}
            \and
            A. L. Fetter \at
            GLAM, McCullough Building\\
            Stanford, CA 94305-4045
            \and
            G. Roati \at
            LENS and Dipartimento di Fisica, Universita di Firenze, and INFM-CNR, Via Nello Carrara 1,\\
            50019 Sesto Fiorentino, Italy
}

\maketitle

\begin{abstract}

In a recent study we demonstrated the emergence of turbulence in a trapped Bose-Einstein condensate of $^{87}$Rb atoms. An intriguing observation in such a system is the behavior of the turbulent cloud during free expansion.The aspect ratio of the cloud size does not change in the way one would expect for an ordinary non-rotating (vortex-free) condensate. Here we show that the anomalous expansion can be understood, at least qualitatively, in terms of the presence of vorticity distributed throughout the cloud,  effectively counteracting the usual reversal of the aspect ratio seen in free time-of-flight expansion of non-rotating condensates.

\keywords{atomic quantum fluid, condensate expansion, vortices, turbulence} \PACS{03.75.Hh \and 05.30.Jp \and 67.40.Db}
\end{abstract}

\section{Introduction}

Atomic quantum fluids have become an important system in the investigation of many effects inherent to superfluidity \cite{Leggett,FootScissors,FootExp,StrinExp,StrinScissors}. One of the most striking properties of a superfluid is the behavior with respect to rotations, which in these systems usually exist in the form of quantized vortices \cite{Pethick-book,Stringbook,Fetter,Fetter2}. Vortices are among the most investigated excitations in superfluids, and since their first appearance in atomic systems \cite{Cornell,Dalibard,Ketterle}, they have also become an import aspect of research in atomic quantum fluids.

Early experimental studies of a single vortex in a large trapped condensate, obtained by stirring techniques, allowed the first investigations of vortex nucleation and stability in these systems \cite{Dalibard2}. Subsequently, the production of a doubly charged vortex state \cite{Shin}, followed by the observation of its splitting into two singly charged vortices, has shown the dynamical instability of these multiply-charged configurations and confirmed that the most energetically favorable configuration contains only singly quantized vortices. As result, in the limit of large rotations and low dissipation, the vortices will equilibrate into an ordered  lattice structure, similar to a crystal.

Vortex lattices have been observed in atomic Bose-Einstein condensates (BEC) by several groups \cite{Ketterle,Dalibard2,Cornell2,Muniz}. Those experiments have shown a broad range of vortex numbers $N_V$, from just a few to over a hundred structures, almost always producing patterns similar to the Abrikosov lattices, observed in type-II superconductors. More recently, vortices have also been produced in trapped quantum gases
by an excitation field that simultaneously combine motion and rotations along multiple axes \cite{Henn09,HennPRL}. Numerical studies using the Gross-Pitaevskii equation (GP) have  shown \cite{Kobayashi-Tsubota07,Tsubota-Kobayashi07} that similar configurations evolve dynamically into the quantum turbulent regime, where  tangled vortices are distributed randomly in the condensate.

Many important aspects of turbulence in quantum fluids have been studied in the seminal work by Vinen \cite{Vinen} and others in superfluid helium. An important motivation to study quantum turbulence has been the understanding of turbulence in a classical fluid, which is a very challenging and important practical problem. The unique restriction of motion of a superfluid, due to its quantum-mechanical constraints, results in a reduced number of degrees of freedom and makes the study of turbulence in quantum fluids (quantum turbulence) an attractive route to gain insight into classical turbulence \cite{Tsubota-QTreview,Vinen-QTreview10,Paoletti-QTreview}. Turbulence in atomic condensates, therefore, creates new exciting possibilities in this important field of research.

In this communication, we present a  simple hydrodynamical model that qualitatively explains the anomalous free expansion observed experimentally in \cite{HennPRL}. Our model describes a condensate with vorticity distributed along different directions, and uses the hydrodynamical description, in the Thomas-Fermi limit, to obtain analytical expressions for the time evolution of the aspect ratio of the expanding cloud of atoms, after its release from the trap.

\section{Production of a turbulent BEC}

In order to study a turbulent superfluid in the laboratory, we first produce a $^{87}$Rb condensate in an axisymmetric  harmonic magnetic trap, with $\omega_{x}=\omega_{y}=\omega_{\rho}=2\pi\times207$ Hz and $\omega_{z}=2\pi\times23$  Hz,  using the procedures described in reference \cite{Henn08}. This produces typically a condensate with $2\times10^{5}$ atoms which is then kept in an elongated (cylindrical) magnetic trap, where an external oscillatory magnetic field is applied, slightly off-axis and for variable amounts of time, by a pair of quadrupole coils. The resulting time-dependent field generates an oscillatory potential which imparts spatial and time varying forces to the atomic cloud. That combination produces simultaneous small rotations, translations and deformations of the condensate.

\begin{figure}[!h]
  \centering
      \includegraphics [width=0.85\textwidth] {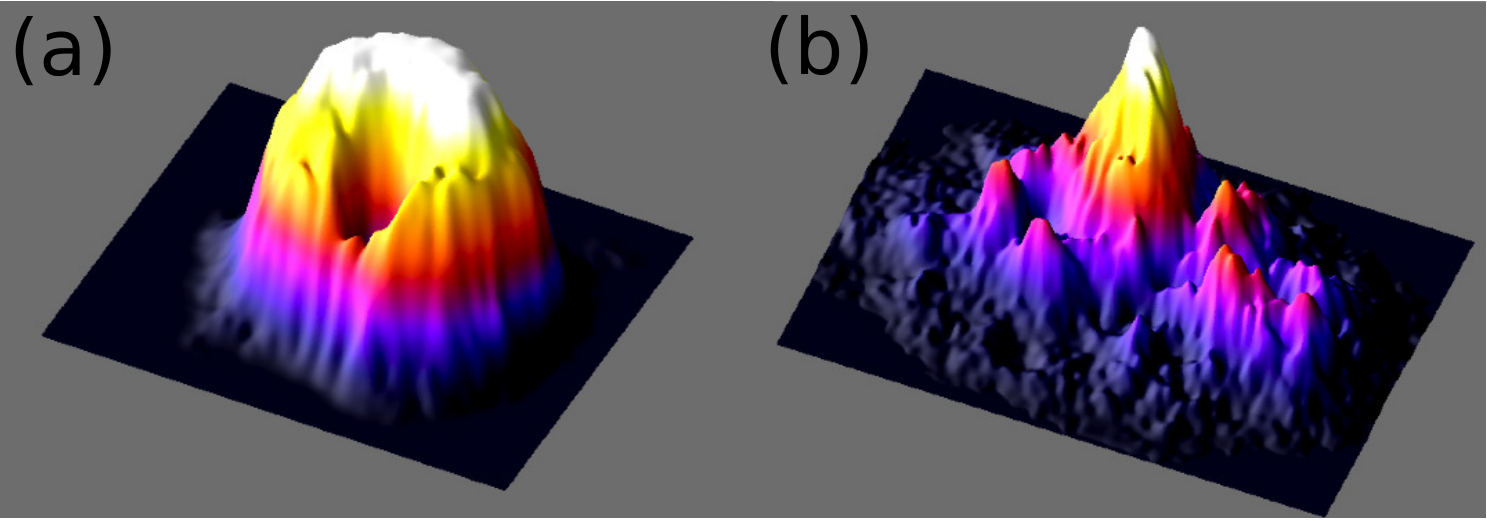}
   \caption{Density profile of the atomic condensate cloud, obtained using absorption imaging after 15 ms of time-of-flight free expansion (images have different scales), showing different regimes: (a) single vortex in an ordinary condensate, where the Field-of-View (FOV) is $60\,\micro\meter \times 80 \,\micro\meter$;  (b) density distribution of a turbulent condensate, with FOV of $300\,\micro\meter \times 100\,\micro\meter$.
\label{fig:fig0na} }
\end{figure}

Within a certain range of parameters, this perturbation leads to the formation of singly quantized vortices in the condensate, as the one present by Fig.~\ref{fig:fig0na}a.
 \noindent
Varying these experimental parameters one can produce very complicated patterns, as seen in Fig.~\ref{fig:fig0na}b, which are reminiscent of the density distribution of a vortex tangle \cite{Kobayashi-Tsubota07,Tsubota-QTreview}. These density fluctuations remain in the cloud for times longer than several tents of milliseconds (longer than it takes for a typical density notch to move across the entire cloud), supporting the idea that these density fluctuations are due to tangled quantized vortices, and not simply a transient sound field caused by the excitation coils. Due to the limitations of the absorption imaging technique, which provides an integrated absorption along the line of sight, one cannot extract the full 3D arrangement of the vortices from images collected along a single direction.  Comparison of the two density profiles  in Fig.~\ref{fig:fig0na}, however, gives a clear distinction between the regular (a) and the turbulent (b)  vortex regimes, based on the higher density fluctuations present in the latter and the significant difference in the cloud sizes. But, as we discuss in the next section, an even more striking effect is the anomalous expansion observed \cite{HennPRL} for a condensate in the turbulent regime. It is this behavior in time-of-flight that motivated the work presented here.

 The exact mechanism of vortex nucleation resulting in the turbulent regime is currently under study. We believe that it may be related to the scheme proposed in \cite{Kobayashi-Tsubota07,Tsubota-Kobayashi07}, where the authors show that numerical simulation of the GP equation reveals turbulence induced in a trapped BEC by combined rotations around two axes (for more details about our trap excitation and the generation of the turbulent regime, see ref.  \cite{Jorge11}).

\section{Observation of a self-similar expansion in a turbulent condensate}

One of the clear signatures of quantum degeneracy in trapped ultracold gases is the asymmetric time-of-flight expansion of the atoms released from an anisotropic trap (see, for example~\cite{Pethick-book}, Chap.~7). This behavior  is very distinct from a classical (thermal) gas, that always expands isotropically (unitary aspect ratio) at long enough time-of-flights. Quantum gases in anisotropic potentials will always have asymmetric velocity distributions, even for the ideal case where the asymmetry comes directly from the uncertainty principle. More often though, as it happens in most experiments involving alkali atoms with repulsive interactions, the effect is enhanced by the non-linear interactions, leading to an excess kinetic energy in the direction of tightest confinement. That happens because of the conversion of the mean-field energy along that direction. This behavior is well established in the field of quantum gases and has been studied and discussed by various authors \cite{Castin-Dum96,StringariRMP}, since the early days of BEC in alkali gases, being the general case typically well described by the GP equation.

Therefore, for an initially ($t=0$) elongated vortex-free condensate with Thomas-Fermi radii $R_\rho(0) \ll R_z(0)$, the expansion is much faster in the radial (tighter) direction.  Consequently, the aspect ratio $R_{\rho}(t)/R_z(t)$ quickly grows in time-of-flight, from its initial small value and reaching an asymptotic value that depends only in the initial aspect ratio and the mean-field energy of the condensate (typically this final value is large). Hence,  the free expansion of a non-rotating condensate produces a characteristic `inversion' of the initial small aspect ratio, becoming larger (or smaller, depending on the directions) than unity at long time-of-flights. In contrast, the asymptotic value of the aspect ratio of a thermal (non-condensed) cloud always tends to the unity.

A remarkable behavior of the turbulent quantum degenerate cloud in free expansion is that it seems to keep the cloud aspect ratio essentially unchanged. In fact, the measurements in ref. \cite{HennPRL} show a complete suppression of the aspect ratio inversion within the range of time-of-flights observed.

\begin{figure}[!h]
  \centering
      \includegraphics [width=0.6\textwidth] {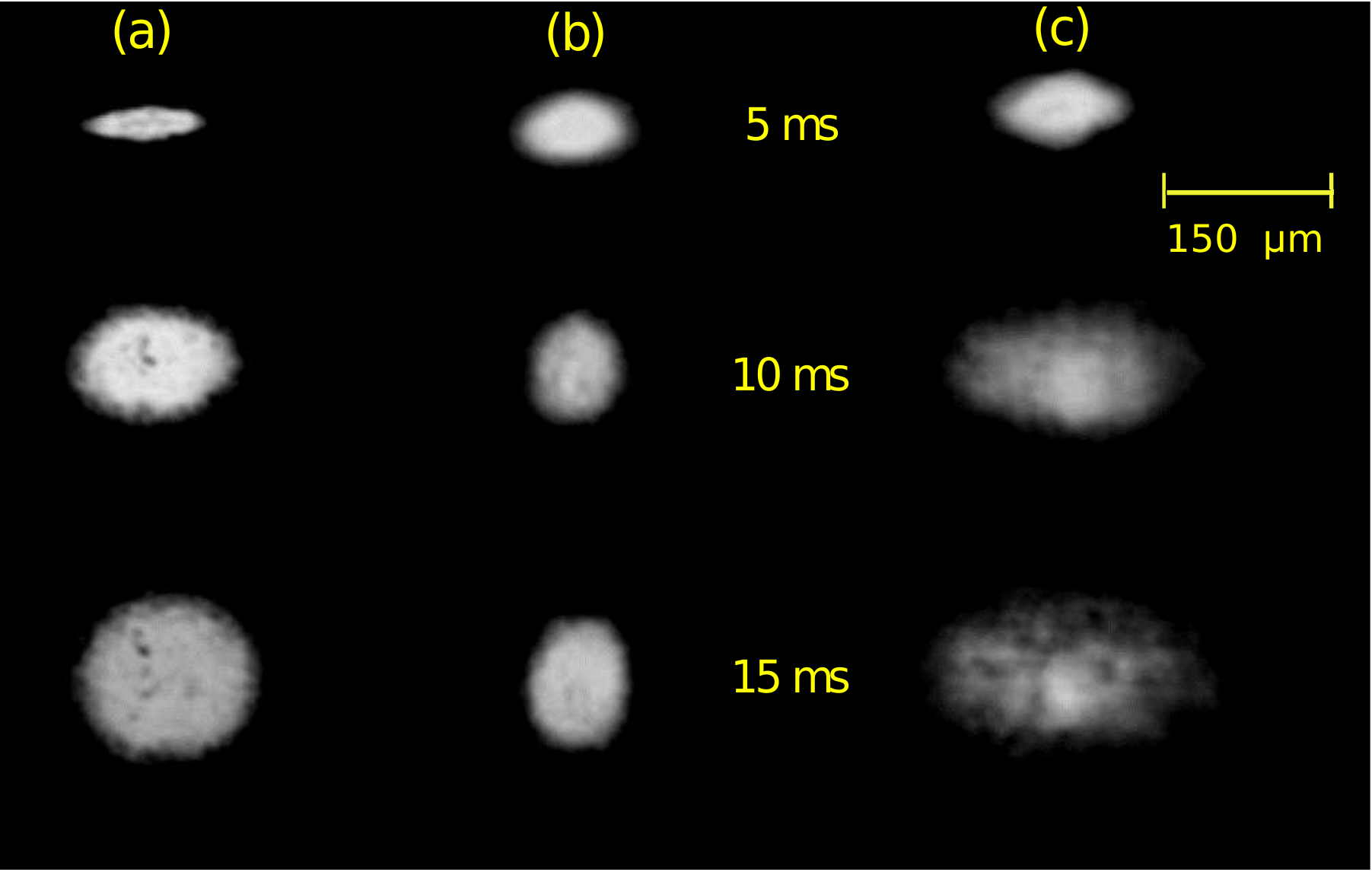}
   \caption{ Evolution of the aspect ratio of the cloud during time-of-flight, for different initial conditions: (a) shows a classical thermal (non-condensed)  cloud expansion, for which the value of aspect ratio tends to unity; (b) an anisotropic expansion of an ordinary BEC, without vortices, where there is an inversion of the initial aspect ratio; (c)  expansion of a condensate in the turbulent regime, where the aspect ratio seems effectively fixed.
\label{fig:ExpansionBEC} } \end{figure}

Here, we call this behavior  self-similar expansion \cite{FetterPrivate,footnote}, and we explain it qualitatively in terms of a distribution of vortex lines, using a very simple model primarily intended to gain some intuition about the much more complicated turbulent regime.  These three regimes are shown in Fig.~\ref{fig:ExpansionBEC}, where one clearly sees that the turbulent cloud expansion (c)  is markedly different from both a non-rotating (vortex-free) BEC (b) and a thermal cloud (a).

\section{Theory}

We start with the hydrodynamic equations (\ref{eq:hdd}) and (\ref{eq:hde}) that  describe the macroscopic dynamics of
the condensate in the same way the GP equation. In these equations, $n$ is the atomic number density, $\overrightarrow{v}$ the velocity field,
$V_{trap}$ the confining harmonic potential, and the interatomic interaction appears in the positive coupling constant  $g=4\pi\hbar^{2}a_{s}/m$
(with $a_{s}$ the s-wave scattering length and $m$ the atomic mass).

\begin{equation}
\label{eq:hdd}
\frac{\partial n}{\partial t}+ \overrightarrow{\nabla}\cdot(n\overrightarrow{v}) =0
\end{equation}
\begin{equation}
\label{eq:hde}
m \frac{\partial \overrightarrow{v}}{\partial t}+
\overrightarrow{\nabla}\left[\frac{1}{2} m v^{2}+V_{trap}+gn \right]=m \
\overrightarrow{v}\times\left(\overrightarrow{\nabla}\times\overrightarrow{v}\right)
\end{equation}

These equations are equivalent to the continuity equation and Euler's equation of a perfect (inviscid) fluid. Typically for a superfluid, they reduce to represent a potential (irrotational) flow, since $\overrightarrow{\nabla}\times\overrightarrow{v}=0$. In that form, these equations are useful to calculate the collective oscillation in  condensates \cite{Pethick-book,Stringbook}, where they have been applied very successfully \cite{DJinPRL,String96PRL,StringariRMP}.

However, in the presence of vortices (singularities) the right-hand side of Eq. (\ref{eq:hde}) is not zero, which significantly alters the relevant physics. For a sufficiently large number of vortices, it is appropriate to write  $\overrightarrow{\nabla}\times\overrightarrow{v}=2\overrightarrow{\Omega} $. This approximation of distributed vorticity  has been used very successfully in Refs. \cite{Sedrakian,SStringari} to describe the dynamics of a condensate containing a vortex lattice. Note that this step  introduces a rotational component in the velocity field of the condensate, such that one can write $\overrightarrow{v}=\overrightarrow{v_{I}}+\overrightarrow{v_{R}} $, with $\overrightarrow{v_{I}}$
representing the irrotational component and $\overrightarrow{v_{R}}=\overrightarrow{\Omega} \times\overrightarrow{r}$. Here $\overrightarrow{\Omega}$ is the macroscopic angular velocity of the sample, and it corresponds to the Feynman expression given in Eq. (\ref{eq:frequency}), with $n_{V}$ being the areal vortex density ($n_{V}= N_{V}/A_{\bot}$), derived when the singular velocity fields of the vortices are distributed uniformly in the plane of rotation $A_{\bot}$. \begin{equation}\label{eq:frequency} \Omega=\frac{h}{2 m}\;n_{V} = \frac{\pi \hbar}{m} \;n_V \end{equation}

The aim here is to describe the free expansion of a BEC containing angular momentum distributed along different directions. To first order we neglect the interactions among vortices, considering only the macroscopic effect of the rotation in each direction. We also do not include any decay mechanism of the vortices (or turbulence) during time-of flight, because of the rapid decrease of the density after release from the trap \cite{Pethick-book,Stringbook}. Starting from the general description of rotational hydrodynamics, we  focus our attention on two simple limiting cases, in order to develop some intuition about the system. In addition, since we are seeking to derive analytical expressions, here we will implicitly use the Thomas-Fermi (TF) approximation, where the kinetic energy is much smaller than the interaction energy.

Consider the following ansatz for the density $n$ and for the velocity field $\overrightarrow{v}$~\cite{SStringari}:
\begin{equation}
\label{eq:density}
n(\overrightarrow{r},t)= n_{0}(t) \left(1-\frac{x^{2}}{R_{x}^{2}(t)}-\frac{y^{2}}{R_{y}^{2}(t)}-\frac{z^{2}}{R_{z}^{2}(t)}\right),
\end{equation}
\begin{equation}
\label{eq:velocity}
\overrightarrow{v}(\overrightarrow{r},t)=\frac{1}{2}\overrightarrow{\nabla}
\left({b}_{x}(t)x^{2}+{b}_{y}(t)y^{2}+{b}_{z}(t)z^{2}\right)+\overrightarrow{\Omega}(t)
\times\overrightarrow{r},
\end{equation}
\begin{equation}
\label{eq:norm}
n_{0}(t)= \frac{15N}{8\pi}\frac{1}{R_{x}(t)R_{y}(t)R_{z}(t)}.
\end{equation}
\noindent Here $R_j^2 =2\mu /m\omega_j^2$ gives the TF radii in terms of the chemical potential $\mu$, and the peak (central) density $n_{0}$  follows from Eq.~(\ref{eq:density}), with the normalization to the total number of atoms $N$. Equations (4) and (6) basically express the TF approximation, known to represent well the system \cite{Pethick-book,Stringbook} under these circumstances, while (5) gives a general description of the velocity field in a rotating condensate. The coefficients $b_{i}(t)$ are temporary unknowns to be determined in what follows.

\begin{figure}[!h]
    \centering
    \includegraphics [angle=0, width=0.9\textwidth] {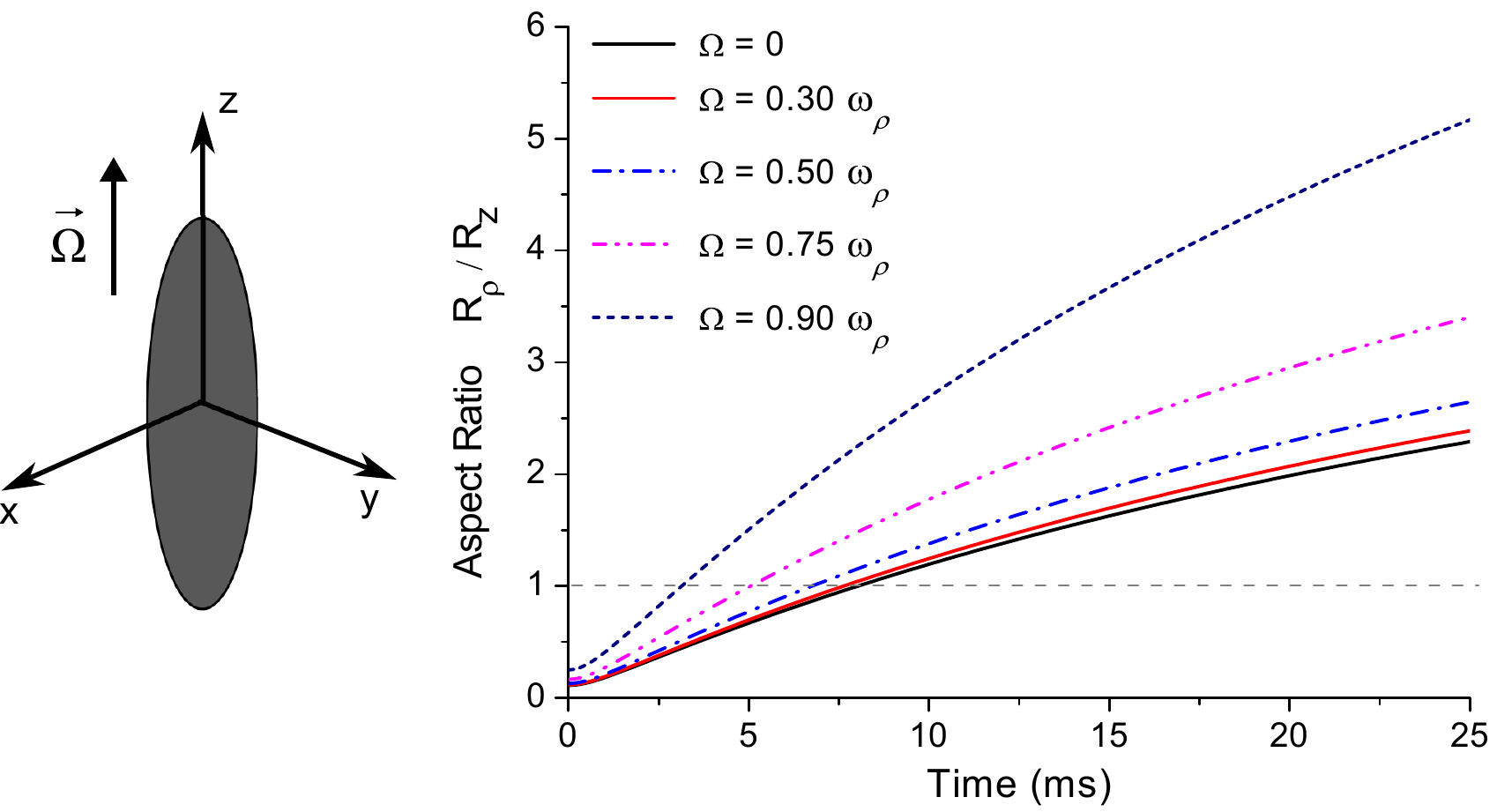}
    \caption{ Representation of the system geometry (left) and evolution of the aspect ratio during time-of-flight (right), given by the Eq. (9), for different values of vorticity in a vortex lattice aligned along the symmetry axis \textit{z}.
    \label{fig:fignsim} }
\end{figure}

As a first example, we consider a vortex array aligned along the symmetry axis, that is, $\Omega$ parallel to $z$ axis (see Fig.~\ref{fig:fignsim}).  Substituting Eq.~(\ref{eq:density}) and Eq.~(\ref{eq:velocity}) in the continuity equation (\ref{eq:hdd}), we find the following relation between the general coefficient in the irrotational velocity ansatz $b_{i}(t)$  and the radius $R_{i}(t)$ :
\begin{equation}\label{eq:bi}
b_{i}(t)=\frac{\dot{R_{i}}(t)}{R_{i}(t)},\,\,_{i=x,y,z} \,.
\end{equation}

Substituting Eq.~(\ref{eq:bi}) in the Euler equation (\ref{eq:hde}), together with the Feynman expression Eq.~(\ref{eq:omegFey}), we find the evolution equations (\ref{eq:evsim}) for the condensate radii ($R_{x} = R_{y} = R_{\rho}$). For clarity, we simplified the notation (dropping the explicit time-dependence) in equation (\ref{eq:evsim}), and used $R_{\rho}$ to represent the radial (transversal) size.
\begin{equation}\label{eq:omegFey}
\Omega(t)=\frac{\hbar}{m} \frac{ N_{V}} {{R_{\rho}(t)}^{2}}
\end{equation}
\begin{eqnarray}
\ddot{R_{\rho}}+\omega_{\rho}^{2}R_{\rho}-\frac{15 N \hbar^{2} a_{s}}{m^{2}}\frac{1}{R_{\rho}^{3}R_{z}}=
\left(\frac{N_{V} \hbar}{m}\right)^2\frac{1}{R_{\rho}^{3}}\nonumber\\
\ddot{R_{z}}+\omega_{z}^{2}R_{z}-\frac{15 N \hbar^{2} a_{s}}{m^{2}}\frac{1}{R_{z}^{2}R_{\rho}^{2}}=0
\label{eq:evsim}
\end{eqnarray}

The numerical simulation of equation (\ref{eq:evsim})  was done with experimental parameters of Ref. \cite{HennPRL}, using a Runge-Kutta $4th$ order method.  Starting from  the equilibrium configuration ($\ddot{R_{i}}(0)=0$), we solved Eq.~(\ref{eq:evsim}) without the trap term to obtain the evolution of the aspect ratio during the free expansion, as shown in Fig.~\ref{fig:fignsim}, for various values of initial vorticity $\Omega$.

It is clear that the extra kinetic energy due to the vorticity produces not only a larger initial aspect ratio, but also an extra acceleration of the radial expansion (perpendicular to $\overrightarrow{\Omega}$). At higher values of $\Omega$ this significantly reduces the time necessary for the inversion of the initial aspect ratio. It is important to note that we constrained the maximum value of $\Omega$ used in our simulations, to maintain the vortex core size ($\xi\propto n_{0}^{-1/2}$)  small compared to the vortex line separation ($\sqrt{\hbar/ m
\Omega}$). Another way to verify the limiting value for $\Omega$ is comparing the effective energy due to the vorticity with the chemical potential: $\mu>\hbar\Omega$. The latter has to be large in order to be consistent with the the TF approximation. Particularly, for the limiting  cases considered here, we have:
\begin{eqnarray*}
N_{V} =&0  &\rightarrow\ \mu = 17 \hbar\bar{\omega},\,\  \hbar\Omega = 0  \\
N_{V} =&70  &\rightarrow\ \mu = 6 \hbar\bar{\omega},\,\  \hbar\Omega = 2 \hbar\bar{\omega}
\end{eqnarray*}
\noindent where both quantities, $\Omega$ and $\mu$, were calculated at $t=0$ with the initial radius $R_{i}(0)$, and they are written in terms of the geometric average trap frequency $\bar{\omega}={({\omega_{\rho}}^2 \omega_{z})}^{1/3}$.

The results of this first configuration illustrate clearly the contribution of the vorticity term, parallel to the symmetry vorticity.  We now move on to consider the case where the vortex lattice is perpendicular to the symmetry axis of the trap (\textit{x-axis}). This is, in fact, the preferred direction for the vortex lines in our experiment and therefore more representative of what we see in the laboratory.

Again we derive equations for the evolution of the condensate radii, subject to this particular situation. Here, due to the asymmetry of the trap potential in the plane perpendicular to the axis of rotation, we included an additional irrotational term to the velocity field ansatz Eq. (\ref{eq:velocity2}) to produce an initially stable configuration~\cite{fetter74}.
\begin{eqnarray}
\overrightarrow{v}(\overrightarrow{r},t)=\frac{1}{2}\overrightarrow{\nabla}
\left( {b}_{x}(t) x^{2}+{b}_{y}(t) y^{2}+{b}_{z}(t) z^{2} \right)+\overrightarrow{\Omega}(t)
\times\overrightarrow{r}+\alpha(t)\overrightarrow{\nabla}(yz)
\label{eq:velocity2}
\end{eqnarray}

Similarly to the previous case, through the continuity equation we find the relations (\ref{eq:bi}), (\ref{eq:omegFey2}) and (\ref{eq:alfa}) for $b_{i}(t)$, $\Omega(t)$ and $\alpha(t)$ respectively:
\begin{eqnarray}
\Omega(t)=\frac{\hbar}{m} \frac{ N_{V}} {R_{y}(t)R_{z}(t)} \,\,,
\label{eq:omegFey2}
\end{eqnarray}
\begin{eqnarray}
\alpha(t)=\frac{R^{2}_{y}(t)-R^{2}_{z}(t)}{R^{2}_{y}(t)+R^{2}_{z}(t)}\,\Omega(t) \,.
\label{eq:alfa}
\end{eqnarray}

\noindent Substituting into  Eq.~(\ref{eq:hde}), we obtain the dynamical equations that describe the free evolution of the radii for this asymmetric configuration Eq.~(\ref{eq:evassim}). The last term on the left-hand side of Eq.(\ref{eq:evassim}) is related to the repulsive interaction force, whereas the term of the right represents the contribution of the vorticity.

\begin{eqnarray}
\ddot{R_{x}}-\frac{15 N \hbar^{2}
a_{s}}{m^{2}}\frac{1}{R_{x}^{2}R_{y}R_{z}}=0 \nonumber\\
\ddot{R_{y}}-\frac{15 N \hbar^{2}
a_{s}}{m^{2}}\frac{1}{R_{y}^{2}R_{x}R_{z}}=4\left(\frac{N_{V} \hbar
}{m}\right)^2\frac{1}{\left(R_{y}^{2}+R_{z}^{2}\right)^2}R_{y}\nonumber\\
\ddot{R_{z}}-\frac{15 N \hbar^{2}
a_{s}}{m^{2}}\frac{1}{R_{z}^{2}R_{x}R_{y}}=4\left(\frac{N_{V} \hbar
}{m}\right)^2\frac{1}{\left(R_{y}^{2}+R_{z}^{2}\right)^2}R_{z}
\label{eq:evassim}
\end{eqnarray}

\begin{figure}[!h]
  \centering
      \includegraphics [width=1\textwidth] {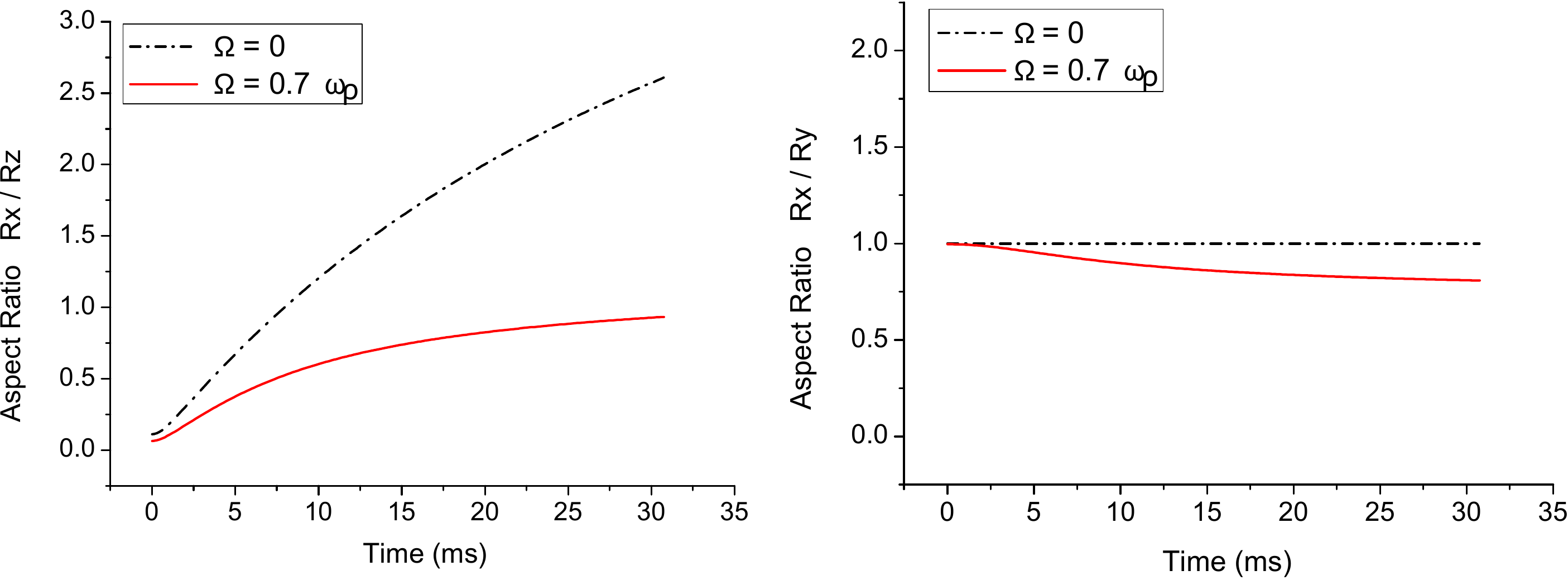}
   \caption{ Evolution of the aspect ratio during time-of-flight, for a system with vorticity along the the \textit{x-axis} (perpendicular to the symmetry axis). The non-inversion of the aspect ratio (the radial divided by the axial dimension) is shown by the solid line of the left hand figure. The figure on the right shows the acceleration of the expansion in the plane perpendicular to the vortex array.
   \label{fig:fignassim3}}
\end{figure}

The corresponding results for the evolution of the aspect ratio are presented in Fig.~\ref{fig:fignassim3}, which summarizes  the most important result of this paper, as it reproduces qualitatively the non-inversion of the aspect ratio, provided the vorticity is above a certain value.
Another point to observe from the simulations is that the vortex  term alters significantly the dynamics of the expansion. The repulsive mean-field interactions act mostly to establish the initial equilibrium configuration and the expansion for  short times (because they involve four inverse powers of the radii $R_i$). In contrast, the centrifugal (vortex)  term with the conserved number of vortices $N_V$ effectively has three inverse powers of $R_{i}$ (the right-hand side of Eq.~(\ref{eq:evassim})).  It  decays more slowly than the repulsive local-density contribution during the free expansion. Hence  the centrifugal term dominates the expansion dynamics at long times.

\section{Conclusions}

In this paper we explain qualitatively the anomalous expansion of a condensate in the turbulent regime. Our model shows that he behavior of the expansion depends intrinsically on the amount and distribution of vorticity present in the sample. To obtain this result we used a simple macroscopic description, based on the rotational hydrodynamics, of a condensate with a relatively large vorticity aligned both parallel and perpendicular to the symmetry axis of the trap. We believe that due to the geometry of the exciting oscillating field in our apparatus \cite{HennPRL}, more angular momentum is coupled into directions perpendicular to the symmetry axis of the trap. That may explain, at least qualitatively, why we obtain such a significant reduction of the aspect ratio inversion in the experiments.

In order to describe quantitatively  the results observed in the laboratory, it is clear that a more elaborate numerical simulation of the turbulent regime, accounting for the distributed vorticity and including the vortex dynamics and possibly reconnection effects (mostly in the trap, since this effect should be small in time-of-flight, when the density decays very rapidly), will be needed to fully account for the experimental observations. The goal here was to present a simple model that seems to capture the main physical mechanism behind the intriguing anomalous free expansion of the cloud.
Although very phenomenological in nature, this simple toy-model already gives a  clear indication that a disordered volumetric distribution of vortex lines can greatly affect the dynamics of the density profile in time-of-flight.

\begin{acknowledgements} We thank M. Tsubota for helpful discussions and E. A. L. Henn, J. A. Seman and D. V. Magalhães for essential assistance in the experiments.\end{acknowledgements}


\begin{thebibliography}{27}
\bibitem{footnote} The term `self-similar' describes  the characteristic expansion (rescalling) of the density profile of the turbulent cloud, which does not change its aspect ratio during the time-of-flight, after releasing from the trap, in contrast to the more usual reversal of aspect ratio for a vortex-free condensate.
\end{thebibliography}
\end{document}